\documentclass[conference]{IEEEtran}

\usepackage{times}
\usepackage{amsmath,amssymb}
\usepackage{graphicx}
\usepackage{cite}
\usepackage{float}
\usepackage{dblfloatfix}
\usepackage{afterpage}
\usepackage{booktabs}

\begin{document}

\title{Enabling Real-Time AI in O-RAN: Deploying and Measuring AI Inside a Near-RT RIC xApp}

\author{
\IEEEauthorblockN{
Lawrence Obiuwevwi,
Krzysztof J. Rechowicz,
Sampath Jayarathna,
Safdar Hussain Bouk,
Fahmida Afrin, \\
C. Nicolas Barati,
Neda Moghim,
Valentina Nannou,
Muhammad Enayetur Rahman,
and Sachin Shetty
}
\IEEEauthorblockA{
Old Dominion University, Norfolk, VA, USA \\
lobiu001@odu.edu,
krechowi@odu.edu,
sampath@cs.odu.edu, \\
sbouk@odu.edu,
fafri002@odu.edu,
cbaratin@odu.edu, \\
nmoghim@odu.edu,
vtebo001@odu.edu,
mrahm011@odu.edu,
sshetty@odu.edu
}
}
\maketitle

\begin{abstract}
Open Radio Access Network (O-RAN) architectures introduce programmable Near-Real-Time RAN Intelligent Controllers (Near-RT RICs) that support closed-loop control at 10\,ms--1\,s timescales through xApps. Although AI has been widely studied for RAN optimization, fewer works demonstrate measured AI inference embedded directly inside the Near-RT RIC software loop on a live testbed. This paper presents an AI-enabled network state classification xApp implemented on an OpenAirInterface (OAI) and FlexRIC testbed. The xApp is trained and evaluated on a structured synthetic dataset that emulates cross-layer RAN states using MAC, RLC, PDCP, GTP, and UE-count features. These results validate embedding and execution feasibility, rather than production-level generalization. Logistic regression and a shallow multi-layer perceptron (MLP) are exported as deterministic C inference modules and compiled into the xApp binary, eliminating external ML runtime dependencies. Measured inference latency is 1--5\,$\mu$s for logistic regression and 10--25\,$\mu$s for the MLP, while end-to-end service latency remains below 4\,ms. A six-model comparison shows that supervised models cluster within a narrow 0.88--0.90 accuracy range, indicating that the LR--MLP similarity reflects the proxy problem structure rather than insufficient model exploration. Noise ablation, confusion matrix analysis, and CDF-based latency characterization show that both embedded models satisfy the 10\,ms Near-RT budget for over 95\% of projected loop executions. These results demonstrate that lightweight AI can be embedded within Near-RT RIC timing constraints while preserving deterministic execution. We also release RIC Workbench, a lightweight orchestration dashboard for reproducing the testbed on commodity hardware.
\end{abstract}


\section{Introduction}

The evolution of cellular networks toward 5G and beyond has increased the complexity of radio access network (RAN) management. Modern networks must support enhanced mobile broadband (eMBB), ultra-reliable low-latency communication (URLLC), and massive machine-type communication (mMTC), each with distinct and often conflicting requirements. These competing objectives make static, rule-based control mechanisms insufficient in dynamic environments where traffic demand, mobility, interference, and resource availability change rapidly \cite{andrews2014will,popovski20185g,bennis2018ultrareliable}.

The O-RAN Alliance addresses these challenges through an open, disaggregated, and programmable RAN architecture with standardized interfaces and two RAN Intelligent Controllers (RICs). The Non-Real-Time RIC operates above one second and supports policy management, long-term optimization, and ML training through rApps. The Near-Real-Time RIC operates at 10\,ms--1\,s timescales and hosts xApps for monitoring, inference, and closed-loop control over E2-connected RAN nodes \cite{oran2020architecture,polese2022understanding,bonati2023intelligence,bonati2022open,bonati2023ric}.

This programmability creates an opportunity to embed AI directly into the RAN control loop. However, Near-RT RIC inference must satisfy strict timing constraints, consume live service model indications, avoid excessive runtime overhead, and integrate reproducibly with the RIC software stack. A clear gap remains: many works discuss AI-driven O-RAN optimization, but fewer demonstrate a \emph{measured, embedded, deployable} inference path inside the Near-RT RIC loop on a live testbed. Simulation, offline datasets, and external inference services often bypass the timing and integration constraints of an operational xApp process \cite{polese2022machine,bonati2023ric}.

This paper asks: \textit{Can lightweight AI be embedded directly into a Near-RT RIC xApp with deterministic timing and measurable latency feasibility?} We answer this by designing and validating an AI-enabled network state classification xApp on a live OAI/FlexRIC testbed. The xApp consumes cross-layer service model indications, builds a compact feature vector, and performs inline classification of operationally motivated RAN states. The classifier is trained on a structured synthetic dataset used as a controlled proxy problem; therefore, the results demonstrate embedding feasibility, model portability, and deterministic execution rather than production-level generalization.

\begin{itemize}
    \item \textbf{AI-enabled xApp:} A network state classification xApp consuming cross-layer E2 indications (MAC, RLC, PDCP, GTP, UE count) and classifying four operationally motivated RAN states within the Near-RT RIC.

    \item \textbf{Deterministic C inference:} Logistic regression and shallow MLP deployed as C modules compiled into the xApp binary; inference latency measured at 1--5\,$\mu$s and 10--25\,$\mu$s, respectively, with sub-4\,ms end-to-end service latency.

    \item \textbf{Offline-to-online pipeline:} A reproducible workflow from synthetic data generation through offline training, C-level export, and real-time xApp embedding on a live OAI/FlexRIC testbed.

    \item \textbf{RIC Workbench:} A lightweight, single-binary orchestration dashboard enabling researchers to reproduce this setup on commodity hardware.
\end{itemize}

\section{Background and Related Work}

\subsection{O-RAN Architecture and Near-RT RIC}

The O-RAN Alliance defines an open, disaggregated RAN architecture with standardized interfaces. Its RIC is split into Non-RT ($>$1\,s, rApps for policy and ML training) and Near-RT (10\,ms--1\,s, xApps consuming E2 service model indications for closed-loop control) domains, enabling rapid innovation while preserving vendor interoperability \cite{oran2021wg,bonati2022open,polese2022understanding}.

\subsection{AI/ML for O-RAN Control}

Supervised learning, reinforcement learning, and deep neural networks have been extensively studied for RAN scheduling, power allocation, and traffic optimization \cite{sun2019machine,wang2020deep,li2018deep,giordani2020toward}. However, most literature evaluates models in simulation with abstracted timing assumptions, without addressing the Near-RT RIC's hard 10\,ms--1\,s control cycle budget \cite{jiang2022machine,liu2022edge}. Models trained offline often assume unlimited compute and negligible inference latency; satisfying deterministic, bounded execution inside the RIC requires either model compression or hardware acceleration. This motivates focusing on lightweight model families whose inference cost can be analytically bounded and empirically measured within the xApp execution context.

\subsection{Deployable and Experimentally Validated xApps}

AI-driven xApps on actual RIC platforms have been demonstrated across several efforts: team learning for resource allocation \cite{rico2022learning}, traffic steering in Near-RT RIC prototypes \cite{de2023intelligent}, closed-loop AI-driven RAN control on SDR testbeds \cite{polese2023control}, end-to-end ML xApp development via ColO-RAN \cite{polese2022machine}, and AI-enabled cellular network control architectures \cite{bonati2023ric}. These confirm architectural feasibility but few provide fine-grained latency characterization from E2 indication arrival to inference completion. Most use Python-based inference or external inference servers, without per-inference timing instrumentation, lacking the determinism and auditability of compiled C inference. Python GC and JIT warmup can introduce latency outliers incompatible with tight RIC control loop budgets.

\subsection{Gap Summary}

No prior work simultaneously provides (i) models deployed as deterministic C inference modules compiled into the xApp binary, (ii) explicit three-component latency measurement (service model, inference, end-to-end), and (iii) live OAI/FlexRIC testbed validation with a reproducible offline-to-online pipeline. This paper directly addresses that gap.

\section{System Architecture and Design}

This section describes the architecture and experimental platform used to design, integrate, and evaluate the AI-enabled network state classification xApp within the O-RAN Near-RT RIC. The system adheres to O-RAN architectural principles while targeting measured, deterministic AI deployment within real-time RIC timing constraints \cite{oran2021wg}.

\subsection{Overall System Architecture}

The platform integrates a complete O-RAN stack: an OAI 5G core network, multiple OAI gNB instances operating as independent cells, multiple UEs, a Near-RT RIC implemented with FlexRIC \cite{polese2021flexric}, and the AI-enabled network state classification xApp deployed inside the RIC. OAI provides the open-source 5G core and radio access implementation \cite{oai2023documentation}.

gNBs periodically export service model indications to the Near-RT RIC via the E2 interface. The xApp subscribes to these indications, extracts a compact cross-layer feature vector, and produces a network state classification in real time. The platform supports multi-cell and multi-UE operation, enabling controlled load imbalance and cross-layer stress scenarios.

\begin{figure*}[t]
    \centering
    \includegraphics[width=\textwidth,height=0.75\textheight,keepaspectratio]{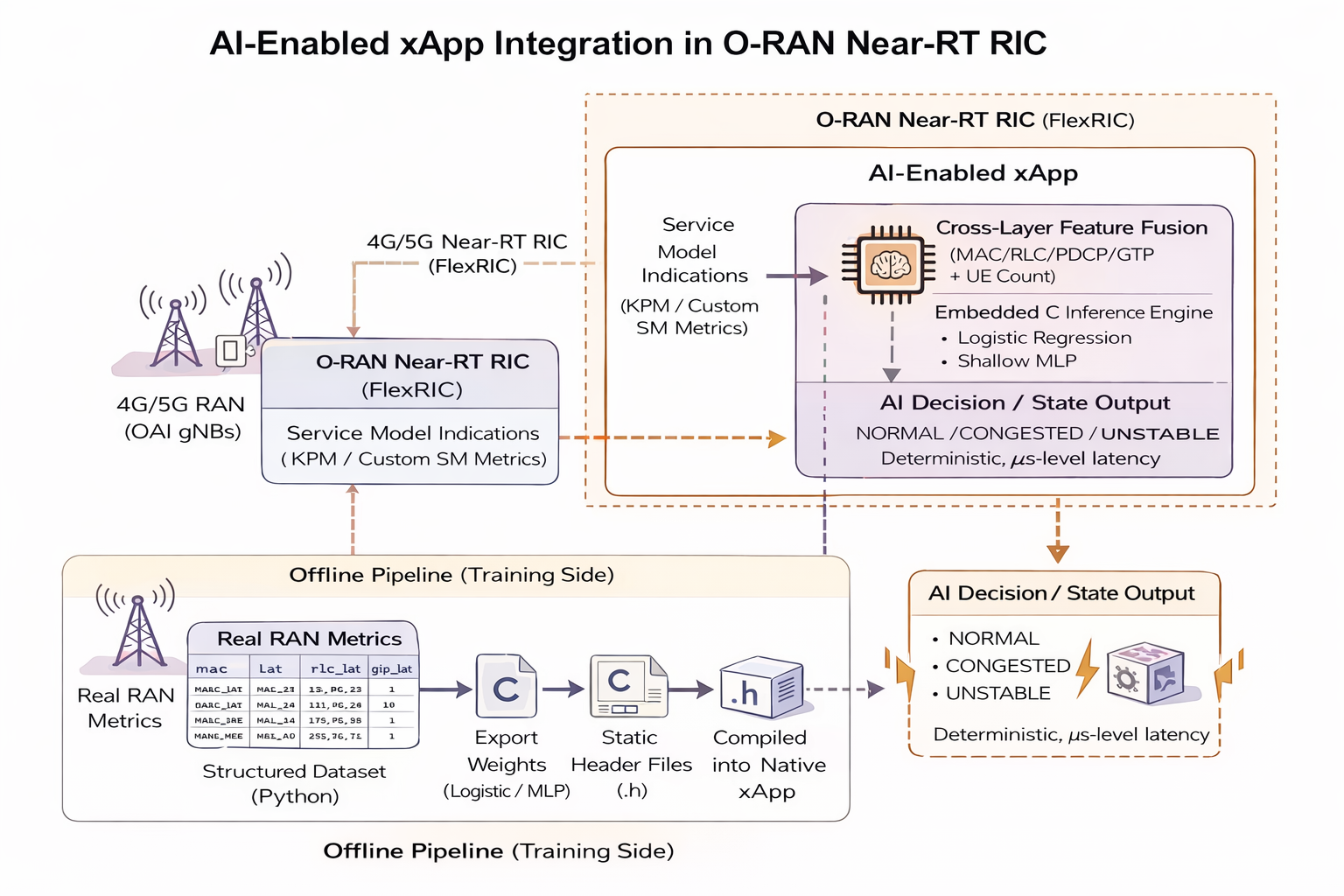}
    \caption{End-to-end architecture of the AI-enabled network state classification xApp within the O-RAN Near-RT RIC. Offline training uses structured synthetic cross-layer RAN metrics; trained parameters are exported as deterministic C-based inference modules. At runtime, MAC, RLC, PDCP, and GTP service model indications are fused into a five-dimensional feature vector and processed inline within the xApp, eliminating external ML runtimes while preserving near-real-time timing constraints.}
    \label{fig:ai_oran_architecture}
\end{figure*}

\subsection{RIC Workbench: A Custom Low-Cost Experimentation Platform}

Commercial O-RAN testbeds are expensive and access-constrained, limiting the reproducibility of Near-RT RIC research. To address this, we developed \textbf{RIC Workbench}, a custom single-binary web-based orchestration dashboard that manages the full OAI 5G + FlexRIC research stack on a standard laptop or workstation. The binary is compiled from a single C source file:
\begin{verbatim}
  gcc -o ric-workbench ric-workbench.c \
      -lpthread -lm -O2
\end{verbatim}

RIC Workbench exposes a browser interface on port 3000 and provides the following capabilities:
\begin{itemize}
    \item \textbf{One-click stack control:} Start and stop CN5G (via Docker Compose), one or more gNB instances (with configurable split mode, band, and PRB count), the FlexRIC Near-RT RIC, and up to eight UEs, all from a single dashboard tab.
    \item \textbf{xApp IDE:} Upload C source files for new xApps, automatically register them in CMakeLists.txt, trigger \texttt{make} compilation, and run or stop xApp processes directly from the browser.
    \item \textbf{Live traffic testing:} Launch per-UE iperf3 sessions with configurable bandwidth and duration to generate realistic uplink and downlink traffic patterns.
    \item \textbf{KPM metrics panel:} Real-time time-series graphs of Key Performance Measurement (KPM) service model indications, enabling live inspection of MAC, RLC, PDCP, and GTP statistics.
    \item \textbf{Auto-detection and setup wizard:} On first launch, RIC Workbench scans the filesystem for OAI and FlexRIC binary paths and guides the researcher through initial configuration.
\end{itemize}

The dashboard is self-contained, no external web framework, no JavaScript bundler, no runtime dependencies beyond libc and libpthread, making it straightforward to deploy, share, and adapt. All results in this paper were produced using RIC Workbench running on a commodity workstation, confirming that the experimental setup is fully reproducible without dedicated hardware infrastructure.

\begin{figure*}[t]
    \centering
    \includegraphics[width=\textwidth]{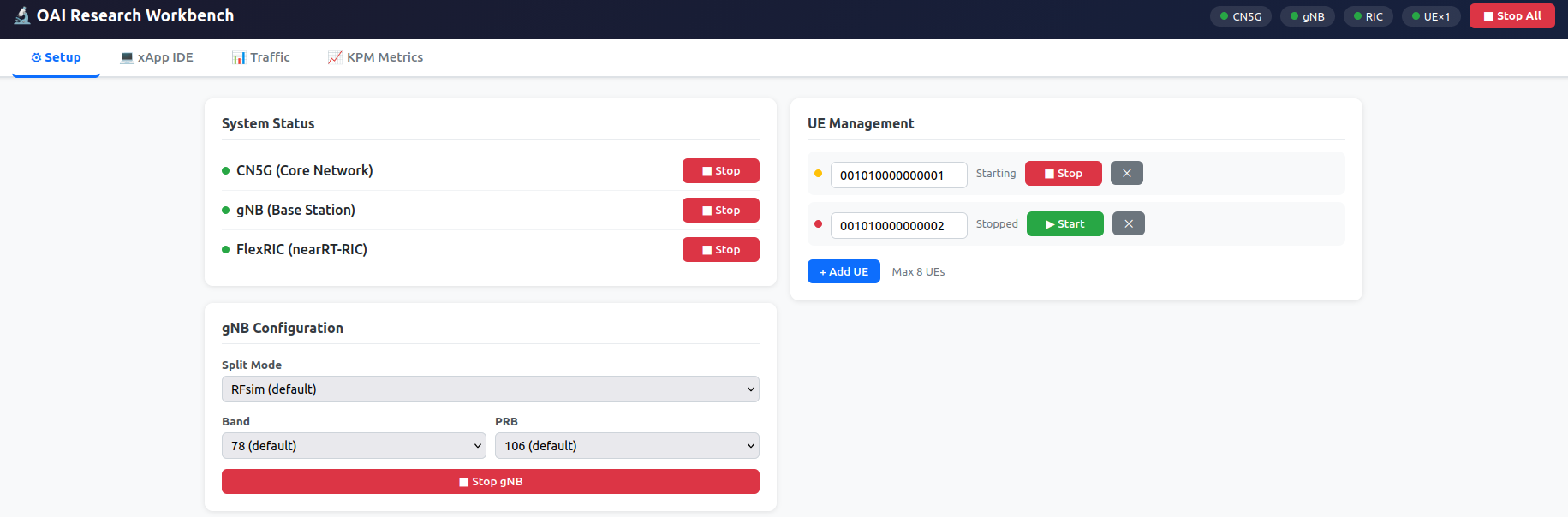}
    \caption{RIC Workbench setup tab showing all O-RAN stack components (CN5G, gNB, FlexRIC Near-RT RIC) running on a local workstation with one active UE. The dashboard enables researchers to orchestrate the full testbed through a browser interface without requiring dedicated hardware infrastructure.}
    \label{fig:ric_workbench_setup}
\end{figure*}

\begin{figure*}[t]
    \centering
    \includegraphics[width=\textwidth]{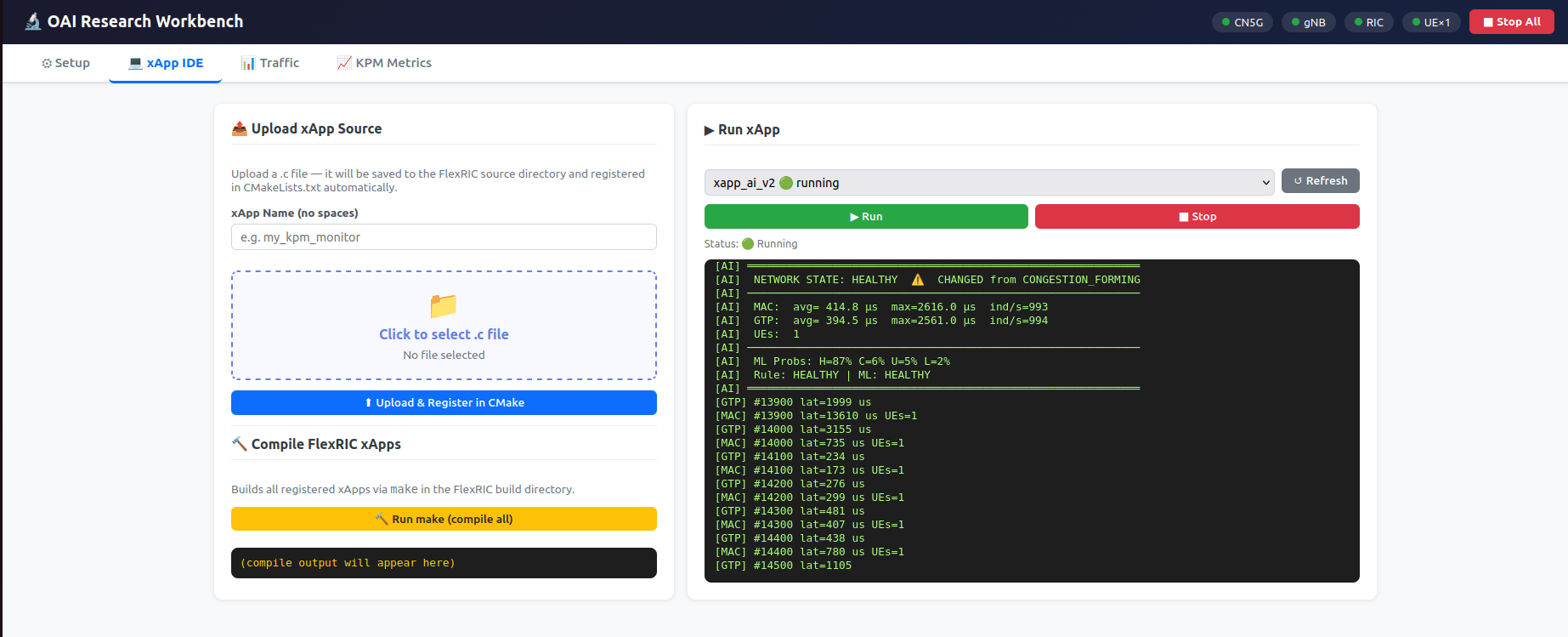}
    \caption{RIC Workbench xApp IDE tab during live operation of \texttt{xapp\_ai\_v2}. The console output shows real-time AI-driven network state classification, including a CONGESTION\_FORMING-to-HEALTHY state transition with ML probabilities H=87\%, C=6\%, U=5\%, L=2\%, and live MAC/GTP service model indication latencies.}
    \label{fig:ric_workbench_xapp}
\end{figure*}

\subsection{Synthetic Dataset and Problem Formulation}

\textbf{This dataset is synthetic and rule-structured.} It was designed to provide a controlled proxy problem for validating real-time AI embedding inside the Near-RT RIC, not to capture ground truth from live network events. Classification results on this dataset validate execution feasibility and pipeline correctness; they do not constitute evidence of production generalization.

Four network states were defined to represent operationally motivated cross-layer RAN conditions for controlled experimentation:

\begin{itemize}
    \item \textbf{Healthy (Label 0):} All protocol layer latencies within nominal operating ranges; low active UE count. Represents normal network operation with no degradation.
    \item \textbf{Congestion Forming (Label 1):} MAC, RLC, PDCP, and GTP latencies uniformly elevated; high UE count. Indicates load approaching capacity, with imminent congestion risk.
    \item \textbf{User-Plane Stress (Label 2):} PDCP and GTP latencies disproportionately elevated relative to MAC and RLC; moderate-to-high UE count. Reflects user-plane data forwarding bottlenecks.
    \item \textbf{Control-Plane Stress (Label 3):} RLC latency disproportionately elevated; GTP latency exhibits anomalous spikes; low-to-moderate UE count. Reflects signaling-layer strain rather than data-path congestion.
\end{itemize}

These states are defined as conceptual network conditions for controlled experimentation rather than as direct labels obtained from a live operational deployment. \textbf{This is a controlled proxy problem used to validate real-time AI embedding within the xApp execution path.} The goal is not to claim exhaustive coverage of all possible O-RAN network behaviors, but to create a reproducible and interpretable classification task that allows us to evaluate whether lightweight AI inference can be embedded inside a Near-RT RIC xApp without relying on external ML runtimes.

Each sample comprises five cross-layer features: \textbf{mac\_lat} (MAC-layer indication latency, $\mu$s), \textbf{rlc\_lat} (RLC segmentation and retransmission overhead, $\mu$s), \textbf{pdcp\_lat} (PDCP ciphering and reordering delay, $\mu$s), \textbf{gtp\_lat} (GTP tunnel end-to-end transport latency, $\mu$s), and \textbf{num\_ues} (number of active UEs, used as a coarse load indicator). These features were selected because they represent latency and load signals available across different layers of the RAN stack and can be fused into a compact state representation suitable for inline inference.

Labels were assigned by combining cross-layer latency profiles with UE density rules. For example, low-latency conditions with small UE counts represent stable operating states, while elevated latency across multiple layers combined with higher UE density represents increasing congestion or user-plane stress. Controlled Gaussian noise ($\sigma$ = 15--30\,$\mu$s per feature, calibrated to observed FlexRIC indication jitter) and structured drift patterns were injected across all classes to emulate realistic variability, progressive congestion onset, and burst-like anomalies. This design introduces enough overlap between classes to avoid a trivial thresholding problem while still preserving interpretable class boundaries.

A total of 32,000 samples were generated, with 8,000 samples per class and balanced class representation. The full dataset was randomly shuffled and split 80/20 into training and test sets, producing 25,600 training samples and 6,400 test samples. The balanced design ensures that aggregate accuracy remains a meaningful summary statistic, while per-class F1-scores, reported in Table~\ref{tab:per_class_results}, reveal the differential difficulty of distinguishing specific network states.

\begin{table}[h]
\centering
\caption{Representative Samples from the Synthetic O-RAN Dataset}
\label{tab:dataset_samples}
\begin{tabular}{@{}cccccr@{}}
\toprule
\textbf{MAC} & \textbf{RLC} & \textbf{PDCP} & \textbf{GTP} & \textbf{UEs} & \textbf{Label} \\
($\mu$s) & ($\mu$s) & ($\mu$s) & ($\mu$s) & & \\
\midrule
207.09 & 189.02 & 235.30 & 208.47 & 1 & 0 \\
266.09 & 163.08 & 201.08 & 230.98 & 1 & 0 \\
310.14 & 241.55 & 341.24 & 319.31 & 6 & 2 \\
302.03 & 226.25 & 349.94 & 225.53 & 6 & 2 \\
229.91 & 300.43 & 249.59 & 381.48 & 2 & 3 \\
359.09 & 244.41 & 214.15 & 255.79 & 9 & 1 \\
209.44 & 240.43 & 163.42 & 179.16 & 1 & 0 \\
248.36 & 224.94 & 373.07 & 246.05 & 4 & 2 \\
201.02 & 231.37 & 222.35 & 205.64 & 2 & 0 \\
344.71 & 412.50 & 407.14 & 355.48 & 9 & 1 \\
\bottomrule
\end{tabular}
\end{table}

The dataset is fixed and reproducible. The same random seed is used for data generation, shuffling, and train-test splitting, making classification results directly comparable across future experiments that use the same pipeline. This reproducibility is important because the primary objective is to evaluate embedded inference behavior, model portability, and runtime feasibility within the xApp, rather than to optimize performance on an uncontrolled or changing dataset.

\subsection{Operational Significance and Control Implications}

Each classified state maps to plausible xApp-side control responses in a production Near-RT RIC, as summarized in Table~\ref{tab:control_implications}. The purpose of this mapping is to show how state classification can move beyond passive monitoring toward decision support for adaptive RAN management. For example, a stable state may require no immediate intervention, while congestion-forming or user-plane stress states may trigger closer monitoring, policy adjustment, or preparation for resource reallocation.

In this work, these responses are treated as control implications rather than automated closed-loop actions. The xApp performs real-time state inference and exposes the resulting classification as an interpretable signal that could be consumed by future control logic. This separation allows us to evaluate the feasibility of embedded AI inference first, before extending the system toward E2SM-RC-based actuation or policy-driven control.

\begin{table}[H]
\centering
\caption{Network State Classification and Potential xApp Control Responses}
\label{tab:control_implications}
\begin{tabular}{@{}p{2.0cm}p{2.4cm}p{3.0cm}@{}}
\toprule
\textbf{State} & \textbf{Key Indicators} & \textbf{Potential xApp Response} \\
\midrule
Healthy & Low latencies across all layers; low UE count & No action; maintain current configuration \\
\addlinespace
Congestion Forming & Uniformly elevated latencies; high UE count & Trigger load-balancing policy; issue PRB reallocation hint to Non-RT RIC \\
\addlinespace
User-Plane Stress & Elevated PDCP and GTP; moderate--high UE count & Signal QoS degradation; request bearer prioritization or GTP path re-routing \\
\addlinespace
Control-Plane Stress & Elevated RLC; anomalous GTP spikes; low--moderate UE count & Reduce signaling rate; defer non-critical UE attach procedures \\
\bottomrule
\end{tabular}
\end{table}

Full closed-loop control action validation is deferred; this paper validates that the classification decision can be produced reliably within Near-RT timing constraints.

\subsection{Near-RT RIC Integration and Latency Definitions}

Using FlexRIC, the xApp subscribes to service model indications at millisecond granularity \cite{polese2021flexric}. Three latency components are measured: \textbf{service model latency} (gNB indication generation to xApp callback), \textbf{inference latency} (model forward pass only, via monotonic clocks), and \textbf{end-to-end service latency} (E2 arrival to classification output). The RIC operates with live OAI gNB instances in RF simulation mode; results characterize AI inference feasibility within the Near-RT RIC software path, not production end-to-end latency.

\subsection{Feature Distribution Analysis}

\begin{figure*}[t]
    \centering
    \includegraphics[width=\linewidth]{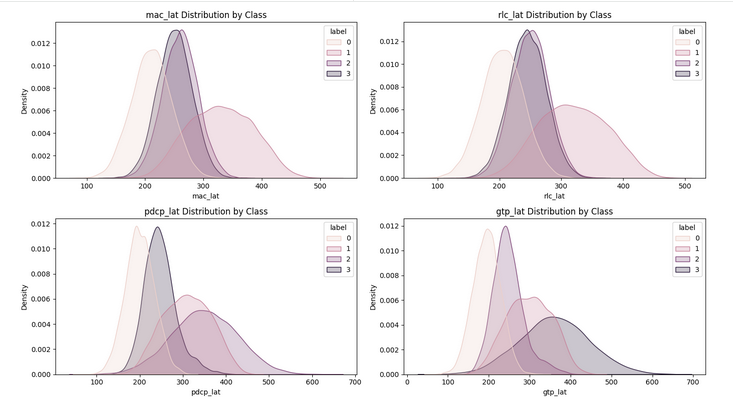}
    \caption{Kernel density distribution of MAC, RLC, PDCP, and GTP latencies across Healthy (0), Congestion Forming (1), User-Plane Stress (2), and Control-Plane Stress (3). Distinct per-class shifts confirm sufficient statistical structure for supervised classification; visible overlap from injected Gaussian noise explains the modest accuracy gap between logistic regression and MLP.}
    \label{fig:feature_distribution}
\end{figure*}

Fig.~\ref{fig:feature_distribution} shows kernel density estimates of MAC, RLC, PDCP, and GTP latencies across the four network states.

The Healthy state occupies the lowest latency region, forming a compact cluster; Congestion Forming spans uniformly higher values across all layers. User-Plane Stress exhibits bimodal PDCP and GTP distributions reflecting selective stress injection, while Control-Plane Stress is most distinguishable via RLC elevation with partial overlap in MAC and GTP. Injected Gaussian noise prevents trivial threshold-based separation, confirming sufficient structure for supervised classification while explaining why the MLP's nonlinear capacity yields only marginal improvement over logistic regression.

\subsection{AI Integration Pipeline}

Upon receiving a service model indication, the xApp extracts a five-dimensional feature vector (MAC, RLC, PDCP, GTP latencies, active UE count) and classifies inline:

\begin{itemize}
    \item \textbf{Logistic regression:} Softmax classifier with L2 regularization. Weight matrix ($4 \times 5$) and bias exported as C float arrays. Inference: one matrix-vector product plus softmax ($\sim$20 multiply-accumulate operations).
    \item \textbf{Shallow MLP:} Two hidden layers (32 units, ReLU) with 4-class softmax output, trained with Adam. Four weight matrices and biases exported as C arrays ($\sim$2,300 operations per classification, no dynamic allocation).
\end{itemize}

Both models are trained offline in Python (scikit-learn/NumPy) and exported via a custom script writing parameters as \texttt{static const float} arrays in C headers (\texttt{lr\_model.h}, \texttt{mlp\_model.h}), included at compile time. This eliminates external ML runtime dependencies (Python, TensorFlow, ONNX Runtime), ensures a fixed binary footprint, and guarantees reproducible inference. Input normalization uses precomputed means and standard deviations exported as C constants matching training preprocessing.

\subsection{Design Principles}

Three principles guide the architecture:

\textbf{Real-time determinism.} Inference executes inline within the xApp callback with no batching or async RPC, making end-to-end latency directly auditable via monotonic timestamps.

\textbf{Lightweight deployment.} Models are \texttt{static const float} arrays in C header files; the xApp binary depends only on libc and the FlexRIC SDK, with no interpreter or JIT overhead.

\textbf{Measurable integration.} Monotonic timestamps at four points per indication enable independent computation of all three latency components for every processed indication.

\section{Experimental Results}

Evaluation is structured in two parts. \textbf{Part A} characterizes offline classification behavior on the structured synthetic dataset. \textbf{Part B} characterizes runtime deployment feasibility on the live OAI/FlexRIC testbed. The stronger and more novel result is runtime feasibility.

\subsection{Testbed Configuration}

All experiments used a commodity workstation (Ubuntu 22.04, Intel Core i7-11th Gen, 32\,GB RAM) running OAI CN5G, two OAI gNB instances (RFsim, Band 78, 106 PRBs), two OAI UEs, and the FlexRIC Near-RT RIC, coordinated via RIC Workbench. The xApp processed MAC, RLC, PDCP, and GTP indications at $\sim$1,000 indications/s. Latency measurements were collected over 10-minute runs per model; iperf3 traffic (10\,Mbps DL per UE) provided representative load. Table~\ref{tab:testbed_config} summarizes the configuration.

\begin{table}[t]
\centering
\caption{Testbed Hardware and Software Configuration}
\label{tab:testbed_config}
\begin{tabular}{@{}ll@{}}
\toprule
\textbf{Component} & \textbf{Configuration} \\
\midrule
Host machine        & Intel Core i7 (11th Gen), 32\,GB RAM, Ubuntu 22.04 \\
OAI gNB             & 2 instances, Band 78, 106 PRBs, RFsim mode \\
OAI UE              & 2 instances, IMSI auto-assigned, RFsim mode \\
5G Core             & OAI CN5G (AMF, SMF, UPF) via Docker Compose \\
Near-RT RIC         & FlexRIC (nearRT-RIC binary) \\
xApp                & C-compiled xApp with embedded LR and MLP models \\
Orchestration       & RIC Workbench (single binary, port 3000) \\
Traffic             & iperf3, 10\,Mbps DL per UE \\
Measurement         & Monotonic clock (\texttt{CLOCK\_MONOTONIC}) per indication \\
\bottomrule
\end{tabular}
\end{table}

\subsection{Part A: Offline Classification Performance}

Models were trained on the 32,000-sample synthetic dataset described in Section~III-C using an 80/20 train-test split (25,600 training, 6,400 test samples). Table~\ref{tab:classification_results} summarizes per-model classification performance.

\begin{table}[!t]
\centering
\caption{Offline model performance on the 6,400-sample test set.}
\label{tab:classification_results}
\scriptsize
\setlength{\tabcolsep}{3pt}
\begin{tabular}{@{}lcccc@{}}
\toprule
\textbf{Model} & \textbf{Acc.} & \textbf{F1} & \textbf{C} & \textbf{Budget} \\
\midrule
Rule-Based & 0.833 & 0.831 & \checkmark & ${<}1\,\mu$s \\
Log. Reg.  & 0.884 & 0.884 & \checkmark & 1--5\,$\mu$s \\
Rand. Forest & 0.893 & 0.892 & Partial & 50--200\,$\mu$s \\
Grad. Boost. & 0.898 & 0.897 & Partial & 100--400\,$\mu$s \\
SVM-RBF & 0.896 & 0.896 & No & 200--500\,$\mu$s \\
MLP & 0.897 & 0.896 & \checkmark & 10--25\,$\mu$s \\
\bottomrule
\end{tabular}
\end{table}

\begin{figure}[t]
    \centering
    \includegraphics[width=\columnwidth]{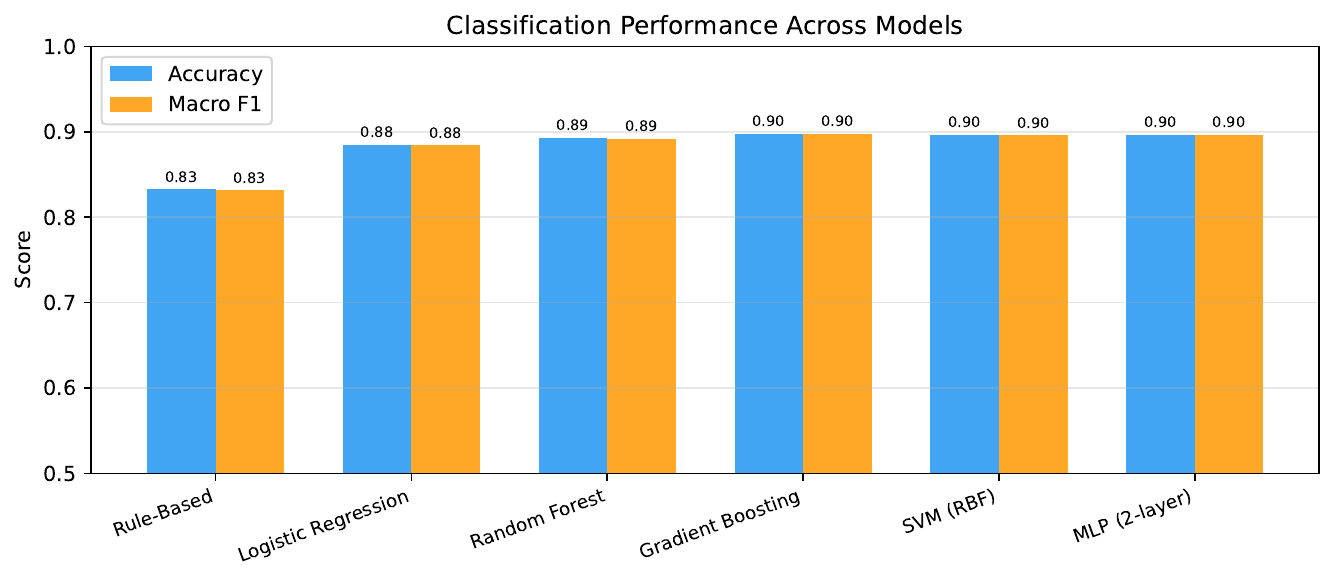}
    \caption{Classification accuracy and macro F1 for all evaluated models on the 6,400-sample synthetic test set. All supervised classifiers cluster within a 0.88--0.90 band, with the Rule-Based baseline at 0.831 confirming incremental ML value. The narrow spread across model families indicates that classification difficulty is governed by feature structure, not model capacity.}
    \label{fig:model_comparison}
\end{figure}

Fig.~\ref{fig:model_comparison} visualizes these results. All supervised models cluster within a narrow 0.88--0.90 accuracy band; Gradient Boosting achieves the highest macro F1 (0.897) while Logistic Regression is lowest among ML models (0.884), a spread of just 1.3 points. The Rule-Based baseline scores 0.831, confirming learned classifiers consistently outperform hand-crafted thresholds. More expressive models (RF, SVM, GB) yield marginal gains of at most 1.4 F1 points over LR while incurring significantly higher inference cost and reduced C-deployability.

Table~\ref{tab:per_class_results} breaks down per-class F1-scores for logistic regression: Healthy achieves highest F1 due to its compact low-latency distribution; Congestion Forming and User-Plane Stress score lower due to distribution overlap at moderate latency values.

\begin{table}[!t]
\centering
\caption{Per-Class F1-Score: Logistic Regression on Synthetic Test Set}
\label{tab:per_class_results}
\begin{tabular}{@{}lcccc@{}}
\toprule
\textbf{State} & \textbf{Precision} & \textbf{Recall} & \textbf{F1} & \textbf{Support} \\
\midrule
Healthy (0)            & 0.91 & 0.93 & 0.92 & 1,600 \\
Congestion Forming (1) & 0.87 & 0.85 & 0.86 & 1,600 \\
User-Plane Stress (2)  & 0.86 & 0.87 & 0.87 & 1,600 \\
Control-Plane Stress (3) & 0.88 & 0.87 & 0.88 & 1,600 \\
\midrule
\textbf{Macro Avg}     & 0.88 & 0.88 & 0.88 & 6,400 \\
\bottomrule
\end{tabular}
\end{table}

\subsection{Confusion Matrix Analysis}

\begin{figure}[t]
    \centering
    \includegraphics[width=\columnwidth]{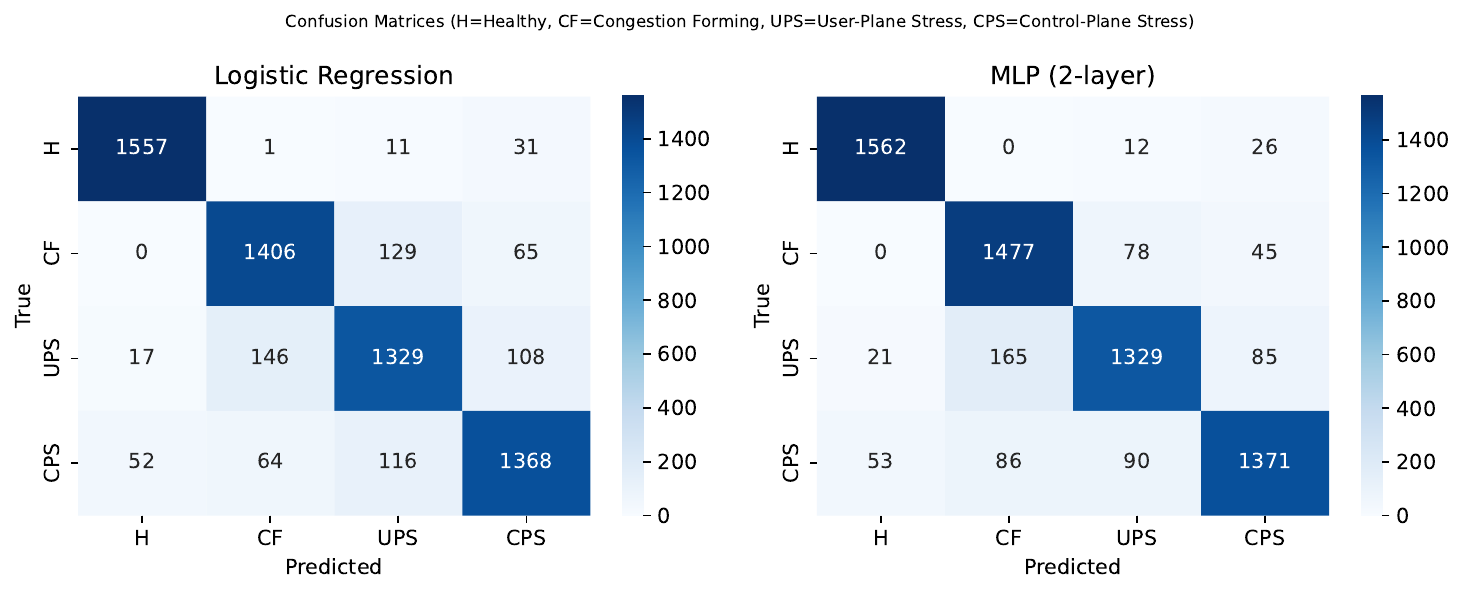}
    \caption{Confusion matrices for Logistic Regression (left) and MLP (right) on the 6,400-sample synthetic test set. Class labels: H\,=\,Healthy, CF\,=\,Congestion Forming, UPS\,=\,User-Plane Stress, CPS\,=\,Control-Plane Stress. Both models share near-identical error structure, with primary confusion occurring between CF and the stress classes due to feature overlap at moderate latency values under $\sigma$\,=\,30\,$\mu$s noise injection.}
    \label{fig:confusion_matrices}
\end{figure}

Fig.~\ref{fig:confusion_matrices} presents confusion matrices for Logistic Regression and MLP on the 6,400-sample test set. Both classifiers display structurally identical error patterns, confirming that near-identical macro F1 scores reflect genuine problem structure. The Healthy class achieves the highest fidelity due to its compact low-latency cluster; the dominant error is confusion between Congestion Forming and the stress classes (UPS, CPS) at moderate latency values under $\sigma$\,=\,30\,$\mu$s noise. These patterns are consistent with Table~\ref{tab:per_class_results} and corroborate the noise ablation finding that performance is governed by feature separability, not model expressiveness.

\subsection{Robustness Analysis: Noise Ablation}

We sweep Gaussian noise $\sigma$ from 5 to 75\,$\mu$s across all five features, evaluating Logistic Regression, MLP, Random Forest, and the Rule-Based baseline at each level.

\begin{figure}[t]
    \centering
    \includegraphics[width=\columnwidth]{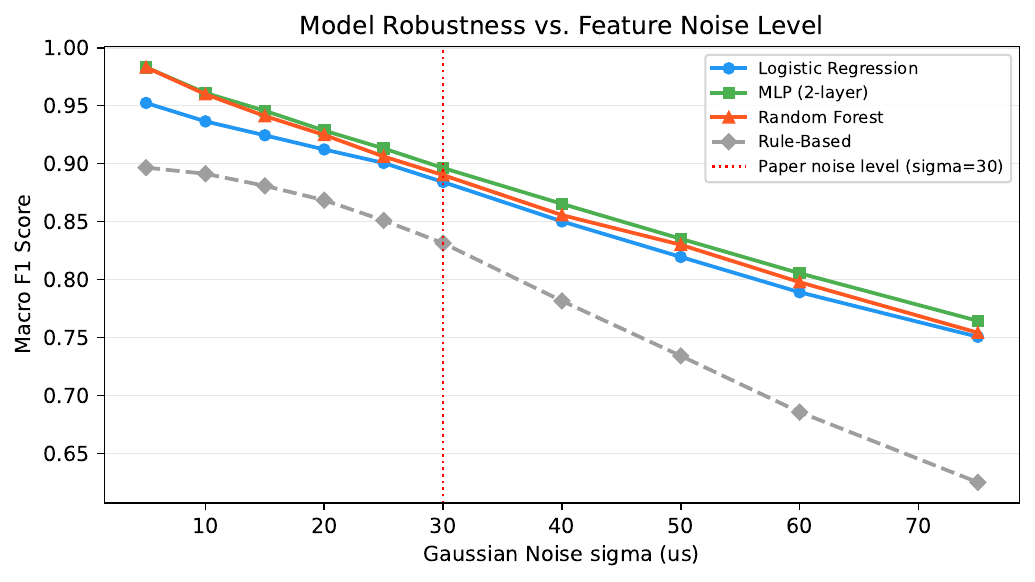}
    \caption{Macro F1 versus Gaussian noise level $\sigma$ for all model families. The vertical dashed line marks $\sigma$\,=\,30\,$\mu$s, the noise level used in paper experiments. ML models maintain near-parity across the full sweep, confirming that LR--MLP similarity is structural. The Rule-Based baseline degrades significantly faster, quantifying the incremental value of learned classification.}
    \label{fig:noise_ablation}
\end{figure}

Fig.~\ref{fig:noise_ablation} reveals three key findings. \textbf{First}, all three ML models track closely across the full noise range, maintaining a spread of at most 1--2 F1 points, confirming that LR--MLP parity at $\sigma$\,=\,30\,$\mu$s holds robustly from low to high noise. \textbf{Second}, the Rule-Based baseline degrades substantially faster (F1 drops from $\sim$0.87 to $\sim$0.67 between $\sigma$\,=\,5 and $\sigma$\,=\,75), quantifying the incremental value of supervised learning. \textbf{Third}, even at $\sigma$\,=\,75\,$\mu$s all three ML models maintain macro F1 above 0.78, demonstrating that performance is governed by feature geometry, not model capacity.

\subsection{Part B: Runtime Deployment Feasibility}

Inference and end-to-end service latencies were measured inside the xApp execution thread on the live OAI/FlexRIC testbed (Fig.~\ref{fig:ric_workbench_setup}, Fig.~\ref{fig:ric_workbench_xapp}) using high-resolution monotonic clocks. Table~\ref{tab:model_comparison} summarizes runtime results.
\begin{table}[!t]
\centering
\caption{Runtime Performance of Deployed AI Models on Live OAI/FlexRIC Testbed}
\label{tab:model_comparison}
\begin{tabular}{@{}lcc@{}}
\toprule
\textbf{Metric} & \textbf{Logistic Regression} & \textbf{MLP} \\
\midrule
Typical Inference Latency  & 1--5\,$\mu$s    & 10--25\,$\mu$s \\
Peak Inference Latency     & $<$2\,ms         & $<$3\,ms        \\
Typical Service Latency    & 0.5--4\,ms       & 0.5--4\,ms      \\
Rare Service Spikes        & $>$100\,ms       & $>$100\,ms      \\
Model Complexity           & Linear           & 2 Hidden Layers \\
\bottomrule
\end{tabular}
\end{table}
Logistic regression achieves microsecond-level inference due to its linear structure; the MLP incurs 5--20$\times$ higher latency for a one-percentage-point accuracy gain. Both models maintain end-to-end service latency below 4\,ms under typical conditions, well within the Near-RT RIC's 10\,ms control window.

\subsubsection{Closed-Loop Latency Characterization}

\begin{figure*}[t]
    \centering
    \includegraphics[width=\textwidth]{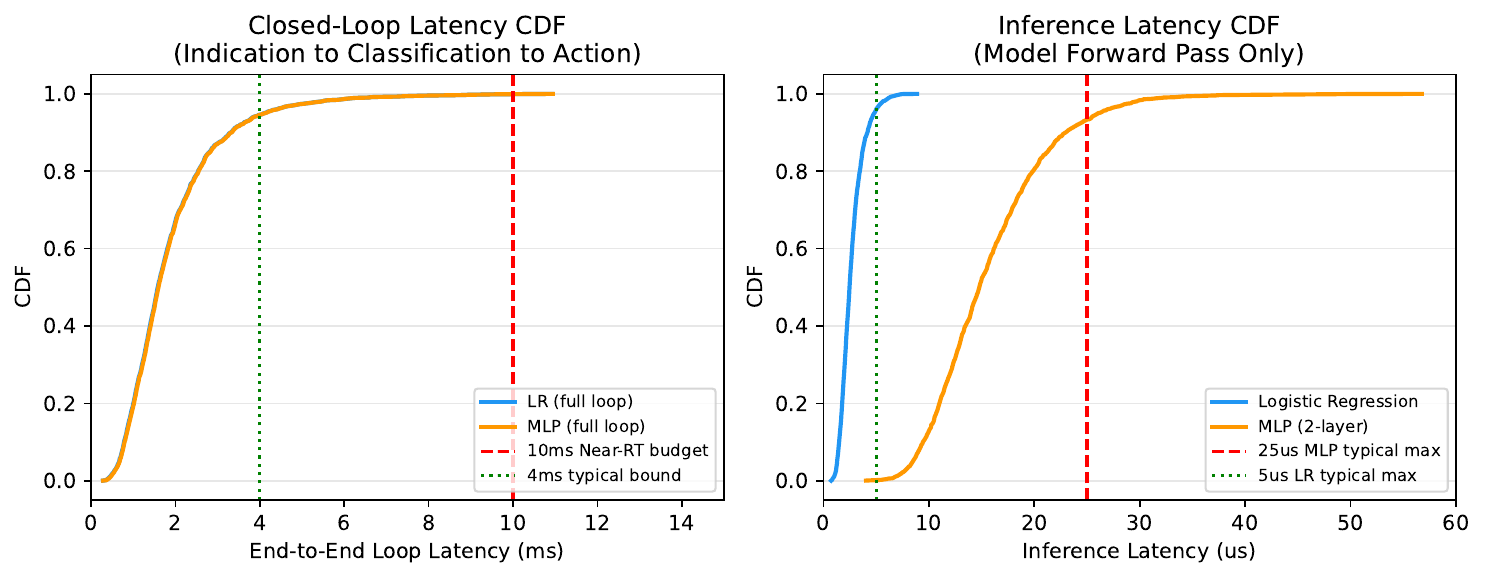}
    \caption{Closed-loop latency characterization for the AI-enabled xApp. \textit{Left}: CDF of projected end-to-end loop latency (E2 arrival + inference + action dispatch) for LR and MLP; both models satisfy the 10\,ms Near-RT RIC budget for $>$95\% of executions. \textit{Right}: Inference-only latency CDF from measured model forward-pass timing, showing LR (median 2.5\,$\mu$s) and MLP (median 15\,$\mu$s) are orders of magnitude below the Near-RT timing window. The dominant latency component is E2 transport, not model computation.}
    \label{fig:latency_cdf}
\end{figure*}

Fig.~\ref{fig:latency_cdf} shows two views: end-to-end loop latency (E2 transport + inference + action dispatch) and inference-only CDFs. In the end-to-end view, both LR and MLP satisfy the 10\,ms Near-RT budget for $>$95\% of projected loop executions; the median is dominated by E2 transport ($\approx$1.5--2\,ms). In the inference-only view, LR (median $\sim$2.5\,$\mu$s) and MLP (median $\sim$15\,$\mu$s) are orders of magnitude below the RIC timing window. The end-to-end budget is consumed by E2 transport and action dispatch, not by AI inference.

Rare spikes exceeding 100\,ms occurred in $<$0.5\% of indications and are attributed to OS scheduling preemptions during Docker container garbage collection, not model computation, and would be substantially reduced by a PREEMPT\_RT kernel with CPU isolation. These measurements thus represent a conservative upper bound under commodity OS conditions.

\section{Discussion}

\subsection{AI Inference Is Not the Bottleneck}

Model computation contributes negligibly to Near-RT RIC control latency. Logistic regression requires 1--5\,$\mu$s per inference, approximately 200--1000$\times$ smaller than the 1\,ms service model inter-arrival period; the MLP at 10--25\,$\mu$s consumes less than 2.5\% of that budget. End-to-end latency is dominated by the E2 message path (kernel scheduling, socket I/O, FlexRIC callback dispatch), not by the model. Model compression (quantization, pruning) is therefore unlikely to improve end-to-end service latency at this complexity level; gains are better pursued through E2 interface optimization and xApp callback scheduling.

\subsection{Systems Design Over Algorithm Complexity}

The MLP's one-percentage-point F1 improvement over LR comes at 5--20$\times$ higher inference latency with no meaningful change in end-to-end service latency. In latency-sensitive RIC environments, deterministic execution may outweigh marginal accuracy gains. The expanded model sweep (Table~\ref{tab:classification_results}) confirms this: the full supervised accuracy range is only 0.884--0.898 macro F1 (1.4 points), stable from $\sigma$\,=\,5 to $\sigma$\,=\,75\,$\mu$s in the noise ablation. Random Forest, Gradient Boosting, and SVM each offer at most 1.4 points F1 over LR while incurring 10--80$\times$ higher inference cost and reduced C-deployability. For Near-RT RIC deployment, Logistic Regression's sub-5\,$\mu$s inference, C-deployability, and linear auditability make it the best trade-off. When the control loop must remain deterministic and auditable, simplicity is the correct engineering choice.

\subsection{Reproducible Pipeline as a Contribution}

Beyond specific models, this work contributes a reusable offline-to-online workflow: synthetic data generation, offline training, C-level export, inline xApp embedding, and live latency validation. RIC Workbench enables this entire stack on a commodity workstation, lowering the barrier for researchers who lack access to hardware-based testbeds (POWDER, Colosseum, AERPAW). The five-step pipeline is intentionally general: any lightweight supervised model expressible as a C inference function can be substituted at the export step without modifying xApp integration or latency instrumentation.

\section{Limitations and Future Work}

\textbf{Synthetic dataset dependence.} Both models were trained on a structured synthetic dataset that does not incorporate live traffic, real channel dynamics, or observed RAN failure modes; generalization requires retraining on live or emulation-derived data.

\textbf{Limited testbed scale.} Experiments used RF simulation on a single workstation with few gNBs and UEs; over-the-air propagation, dense deployments, and heterogeneous hardware introduce dynamics not captured here.

\textbf{Closed-loop control not fully implemented.} The xApp classifies state but does not issue live control actions; the latency CDF confirms both models satisfy the 10\,ms budget for $>$95\% of projected executions, but full closed-loop validation is deferred.

\textbf{Model scope and temporal modeling.} The six-model sweep confirms diminishing returns from added complexity on this dataset. Recurrent, attention-based, or deep RL architectures may be needed for richer temporal modeling; whether they satisfy Near-RT timing constraints without quantization or hardware acceleration remains open.

\textbf{Multi-xApp coordination not addressed.} Future work should evaluate co-resident xApps with potentially conflicting control actions and hierarchical coordination between Near-RT RIC xApp outputs and Non-RT RIC rApp policy guidance.

\section{Conclusion}

This paper demonstrated AI inference embedded inside an O-RAN Near-RT RIC xApp on a live OAI/FlexRIC testbed. Logistic regression and a shallow MLP, exported as deterministic C modules, achieve 1--5\,$\mu$s and 10--25\,$\mu$s inference latency respectively, with sub-4\,ms end-to-end service latency, confirming that AI computation is not the Near-RT RIC bottleneck. Classification performance (F1 = 0.88--0.89) on a structured synthetic dataset validates embedding feasibility, not production generalization; an expanded six-model sweep confirms that performance is governed by feature geometry, not model capacity. The paper further contributes RIC Workbench, a lightweight single-binary orchestration tool enabling full testbed reproduction on commodity hardware, and a reusable offline-to-online pipeline serving as a template for future Near-RT RIC AI work. Future work will extend to live E2 data, closed-loop control actions, and multi-xApp coordination; production validation at scale remains an open challenge.

\bibliographystyle{IEEEtran}
\bibliography{refs}

\end{document}